\documentclass[12pt]{article}
\usepackage{graphics}
\usepackage{amssymb}
\usepackage{amsmath}

\newcommand{\rf}[1]{(\ref{#1})}
\newcommand{\beq}{\begin{equation}}
\newcommand{\eeq}{\end{equation}}
\newcommand{\bea}{\begin{eqnarray}}
\newcommand{\eea}{\end{eqnarray}}

\newcommand{\e}{\mbox{e}}
\renewcommand{\d}{\mbox{d}}

\renewcommand{\l}{\lambda}

\renewcommand{\a}{\alpha}

\newcommand{\del}{\delta}

\newcommand{\oh}{\frac{1}{2}}

\newcommand{\dg}{\dagger}

\newcommand{\tr}{\mathrm{tr}\,}
\newcommand{\ra}{\rangle}
\newcommand{\la}{\langle}
\newcommand{\prt}{\partial}
\newcommand{\mi}{\!-\!}

\newcommand{\pl}{\!+\!}

\newcommand{\cD}{{\cal D}}

\newcommand{\cN}{{\cal N}}

\newcommand{\hH}{{\hat{H}}}

\newcommand{\sla}{\sqrt{\l}}

\newcommand{\vac}{|0\ra}
\newcommand{\cav}{\la 0 |}
\newcommand{\dll}{\frac{dl}{l}}

\begin{document}

\begin{center}

{ \Large \bf A Matrix Model for
2D Quantum Gravity}\\ 
\vspace{10pt}
{\Large\bf 
defined by Causal Dynamical Triangulations }

\vspace{30pt}

{\sl J.\ Ambj\o rn}$\,^{a,b}$, {\sl R.\ Loll}$\,^{b}$,
{\sl Y.\ Watabiki}$\,^{c}$, {\sl W.\ Westra}$\,^{d}$ and
{\sl S.\ Zohren}$\,^{e}$

\vspace{24pt}

{\footnotesize

$^a$~The Niels Bohr Institute, Copenhagen University\\
Blegdamsvej 17, DK-2100 Copenhagen \O , Denmark.\\
{ email: ambjorn@nbi.dk}\\

\vspace{10pt}

$^b$~Institute for Theoretical Physics, Utrecht University, \\
Leuvenlaan 4, NL-3584 CE Utrecht, The Netherlands.\\
{ email: loll@phys.uu.nl}\\

\vspace{10pt}

$^c$~Tokyo Institute of Technology,\\ 
Dept. of Physics, High Energy Theory Group,\\ 
2-12-1 Oh-okayama, Meguro-ku, Tokyo 152-8551, Japan\\
{email: watabiki@th.phys.titech.ac.jp}\\

\vspace{10pt}

$^d$~Department of Physics, University of Iceland,\\
Dunhaga 3, 107 Reykjavik, Iceland\\
{ email: wwestra@raunvis.hi.is}\\

\vspace{10pt}

$^e$~Blackett Laboratory, Imperial College,\\
London SW7 2AZ, UK, and\\
{email: stefan.zohren@imperial.ac.uk}

}
\vspace{48pt}

\end{center}


\begin{center}
{\bf Abstract}
\end{center}

\noindent A novel continuum theory of two-dimensional quantum gravity, based on a version of
Causal Dynamical Triangulations which incorporates topology change, 
has recently been formulated as a genuine
string field theory in zero-dimensional target space (arXiv:0802.0719). 
Here we show that the Dyson-Schwinger equations of this 
string field theory are reproduced by a cubic matrix model. 
This matrix model also appears in the so-called
Dijkgraaf-Vafa correspondence if the superpotential there is required to 
be renormalizable. 
In the spirit of this model, as well as the original large-$N$ expansion by
't Hooft, we need no special double-scaling limit involving a fine tuning
of coupling constants to obtain the continuum quantum-gravitational theory. Our
result also implies a matrix model representation of the original, strictly causal
quantum gravity model.

\vspace{12pt}
\noindent


\newpage

\section{Introduction}

Dynamical triangulations (DT) were introduced as a regularization
of the Polyakov bosonic string and of two-dimensional quantum 
gravity \cite{ambjorn,david,mkk}.
Using this regularization, one could show that a tachyon-free
version of Polyakov's bosonic string theory does 
not exist in target space dimensions 
$d > 1$ \cite{ad}. However, when viewed as a theory of 2d quantum 
gravity coupled to matter with central charge $c\leq 1$, 
the theory (non-critical string theory) did 
make sense. Using matrix-model techniques
and other combinatorial methods, it was sometimes
even advantageous to use the regularized theory for analytic 
calculations. Related attempts to use DT as a regularization of
higher-dimensional quantum gravity \cite{aj} were less successful 
\cite{bielefeld}. This triggered the introduction of 
Causal Dynamical Triangulations (CDT), which use causal,
Lorentzian instead of Euclidean curved spacetimes 
as a fundamental input. Evidence has been accumulating that they
provide us with a non-trivial
theory of quantum gravity in four dimensions \cite{agjl,ajl}.

While the higher-dimensional DT and CDT theories of quantum gravity 
at this point rely strongly on numerical simulations, the 2d CDT theory of quantum gravity 
can be solved analytically \cite{al}, like its 2d Euclidean DT counterpart.
This is described in detail in two recent papers, where we have also 
developed a complete string field theory in a zero-dimensional target
space for the CDT version of 
2d quantum gravity \cite{alwz,alwwz}.\footnote{Due to the inclusion of 
higher-genus surfaces, this amounts to a non-trivial generalization of
the original, strictly causal CDT formulation.}  
This string field theory or third quantization
of 2d quantum gravity uses the formalism already developed 
by Ishibashi, Kawai and collaborators for the DT version of 2d quantum gravity
in the context of non-critical string theory \cite{sft,moresft}. 
For non-critical string theory, it is known from \cite{sft} that 
the string field theory reproduces the results of the {\it double-scaling limit} 
of the matrix models whenever the results can be compared. 

Given the formal similarity between the CDT string field theory and the non-critical
string field theory, it is natural to ask whether there also exists a matrix model which
reproduces the results of the former. Below we will show that 
the answer is in the affirmative. However, since the scaling found in
the CDT model is different from the conventional double-scaling 
limit of matrix models, a different limit needs to be taken.
We will show that the limit is simply the conventional 
limit used in the context of the Dijkgraaf-Vafa duality
to $U(N)$ supersymmetric gauge theories \cite{dv}.

\section{CDT string field theory}

We have recently developed a string field theory for Causal Dynamical 
Triangulations \cite{alwwz}. The starting 
point of the CDT quantization of gravity is the assumption 
that in a gravitational path integral over spacetimes with a Lorentzian signature
only causal geometries should be included, an idea dating 
back at least to \cite{tei}. How this can be done 
explicitly in a regularized theory, how one can rotate to 
Euclidean signature to perform explicit calculations, and eventually take 
the cut-off (or lattice 
spacing) to zero is described in detail in \cite{al} for two and 
in \cite{blp} for three and four spacetime dimensions. We demonstrated in 
\cite{alwz} how one can still solve the 2d model analytically when
the original formulation is extended to allow the light-cone structure to become  
degenerate in isolated points. In \cite{alwwz} we generalized these results
to a genuine string field theory, which enabled us in principle to calculate the 
amplitudes of certain spatial correlators, for two-dimensional worldsheets of
any topology\footnote{For earlier results in this direction 
we refer to \cite{lw}.}. 

Let us briefly define the CDT string field theory, while referring to \cite{alwwz} for details.
We will work in a Euclidean notation, which means that we started out
with a Lorentzian signature, regularized the theory, rotated it to 
Euclidean signature as described in \cite{al} and then took the 
lattice cut-off $a$ to zero. In particular, this implies that all quantities discussed below
are already {\it continuum} quantities. 

We have a ``free'' Hamiltonian $H_0$ which describes  
the causal propagation of a spatial universe with respect 
to proper time $t$. Let a spatial universe with the topology of a
circle of length $l_2$ (the ``exit''
loop) be separated a geodesic distance $t$ from another
spatial loop of length $l_1$ (the ``entrance'' loop), and denote the corresponding amplitude by
$G_\l^{(0)}(l_1,l_2;t)$. It is represented by the path integral
\beq\label{1.0}
G_\l^{(0)} (l_1,l_2;t) =
\int \cD [g_{\mu\nu}] \; e^{-S[g_{\mu\nu}]},
\eeq  
with the (Euclidean) gravity action 
\beq\label{1.1}
S[g_{\mu\nu}] = \l \int \d^2 \xi  \sqrt{\det g_{\mu\nu}(\xi)} +
x \oint \d l_1 + y\oint \d l_2,
\eeq
where $\l$ is  the cosmological constant, 
$x$ and $y$ are two so-called boundary
cosmological constants, $g_{\mu\nu}$ is a metric representing the geometry
(diffeomorphism equivalence class) $[g_{\mu\nu}]$,
which is assumed to be strictly causal in the sense
of the original CDT model \cite{al}. This means essentially that
the topology of its spatial sections will not change as time advances. 
We choose the spacetime to have 
the topology of a cylinder, $S^1\times [0,1]$. In a Hilbert space language one has \cite{al}
\beq\label{1}
G_\l^{(0)}(l_1,l_2;t) = \la l_2 | \e^{-t \, H_0(l)} | l_1\ra, 
 ~~~~
H_0(l) = -l \frac{d^2}{d l^2} +\l l.
\eeq
Next, we will generalize the class of geometries integrated over
in the path integral \rf{1.0}. As a
function of time $t$, spatial geometries will be allowed to
branch into disconnected circles, and the resulting baby universes
can subsequently merge again. Furthermore, a spatial universe will be 
allowed to vanish into the ``vacuum'' if it has length zero. 
These topology-changing processes 
can be described by the string field Hamiltonian
\bea\label{s8}
\hH &=& \int \dll\; \Psi^\dg(l) H_0(l)\Psi(l) - 
g \int dl_1 \int dl_2 \Psi^\dg(l_1)\Psi^\dg(l_2)\Psi(l_1+l_2)
\\ && - \a g\int dl_1 \int dl_2 \Psi^\dg(l_1+l_2)\Psi(l_2)\Psi(l_1)
-\int \dll \; \del(l) \Psi(l). \nonumber
\eea
The operator $\hH$ is a ``second quantized'' Hamiltonian in the sense of 
many-body theory. We introduce creation and annihilation 
operators $\Psi^\dg(l)$ and $\Psi (l)$
for universes of length $l$ which act on the above-mentioned
vacuum state $\vac$, with defining relations  
\beq\label{s1} 
| l\ra = \Psi^\dg (l) \vac,\;
\Psi (l) | l\ra = \vac,\;
\Psi(l)\vac = \cav \Psi^\dg(l) =0,\;\;\;
[\Psi(l),\Psi^\dg(l')]=l\del(l-l'). 
\eeq
In \rf{s8}, $g$ is a coupling constant of mass dimension 3, and $\a$ is a dimensionless
parameter allowing us to distinguish between the merging and splitting
of universes, which will be set to 1 at the end of the calculation.
For $\a=1$, $\hH$ is hermitian except for the presence of the tadpole term
proportional to $\del (l)$.
It tells us that a universe can vanish when it has 
zero length, but cannot be created from nothing.
Also the two interaction terms have a straightforward geometric interpretation.
The first term
replaces a single spatial universe of length $l_1+l_2$ with two
spatial universes of length $l_1$ and $l_2$, while the second term represents
the time-reversed process where two spatial universes merge into one, again without
changing the total length $l_1+l_2$. 
The coupling constant $g$ clearly takes on the role
of string coupling constant, since 
the splitting of spatial universes is associated with a factor $g$ and the
merging with a factor $\a g$, making for a combined factor of 
$\a g^2$ whenever the spacetime topology changes (see \cite{alwwz}
for a detailed discussion).

We can use the string field theory associated with $\hH$ to calculate 
connected multi-loop correlators defined by
\beq\label{3}
w(l_1,\ldots,l_n) = \lim_{t\to \infty} \;\cav \;\e^{-t \hH} 
\Psi^\dg(l_1) \cdots \Psi^\dg (l_n) \;\vac_{connected}. 
\eeq
\begin{figure}[t]{
\centerline{\scalebox{0.5}{\rotatebox{0}{\includegraphics{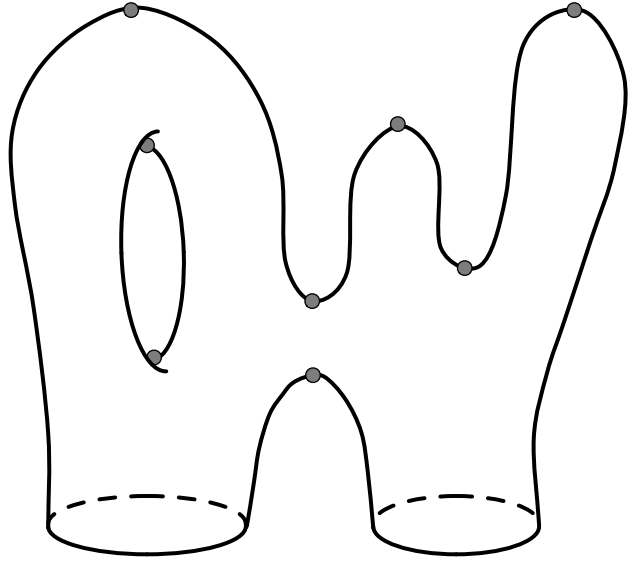}}}}
\caption[fig2]{{\small
A typical geometry in the string field theory contributing to
the amplitude $w(l_1,\ldots,l_n)$ of eq.\ \rf{3}. Proper time progresses 
upwards. The dots mark singular points of the causal structure. }}
\label{fig1}
}
\end{figure}
They describe all possible ways in which an initial state of $n$ spatial loops can evolve
and eventually vanish into the vacuum, while forming a connected 
two-dimensional geometry (c.f. Fig.\ \ref{fig1}).
The amplitudes $w(l_1,\ldots,l_n)$ 
are determined from the string field theory ``partition function''
\beq\label{4}
Z(J) = \lim_{t\to \infty}\; \cav \; \e^{-t\hH} \, \e^{\int \mathrm{d} l 
J(l) \Psi^\dg (l)}\vac
\eeq
through the prescription
\beq\label{5}
w(l_1,\ldots,l_n) = 
\left.\frac{\del^n F(J)}{\del J(l_1) \cdots \del J(l_n)}\right|_{J=0},~~~~
 F(J) = \log Z(J).
\eeq
In \cite{alwwz} we derived the Dyson-Schwinger equations
for the correlators $w(l_1,\ldots,l_n)$. They follow from the
$t$-independence of $Z(J)$, which leads to the relation
\bea
0= \int_0^\infty dl \, J(l)
\left\{  H_0(l)\, \frac{\del F(J)}{\del J(l)} - \delta(l) 
 -g l \int_0^l dl'\;\frac{\del^2 F(J)}{\del J(l')\del J(l-l')} \right. 
\nonumber\\
\left. -g l \int_0^l dl'
\frac{\del F(J)}{\del J(l')}\frac{\del F(J)}{\del J(l-l')}
- \a g l\int_0^\infty dl' l'J(l')\frac{\del F(J)}{\del J(l+l')}\right\}.
\label{ds9}
\eea
One obtains the Dyson-Schwinger equations for the amplitudes 
$w(l_1,\ldots,l_n)$ by differentiating 
\rf{ds9} $n$ times with respect to $J(l)$ and then setting $J(l)=0$. 
The general equation at order $n$ can be written down easily, but is
involved. We will give only the first three equations explicitly, from which the
general structure should be clear.
The Dyson-Schwinger equations are most conveniently formulated in terms of the 
Laplace-transformed amplitudes
\beq\label{ds11}
w(x_1,\ldots,x_n) \equiv \frac{1}{\a^{n-1}} \int_0^\infty dl_1
\cdots\int_0^\infty dl_n \; \e^{-x_1l_1-\cdots -x_nl_n} 
w(l_1,\ldots,l_n),
\eeq
where for convenience we have rescaled the amplitudes 
by a factor $\a^{n-1}$, compared with the convention used in \cite{alwwz}. 
Introducing the notation
\beq\label{V}
V'(x) = \frac{1}{g} (\l -x^2),~~~~
V(x)= \frac{1}{g} \Big(\l x -\frac{1}{3} x^3\Big),
\eeq
we obtain from \rf{ds9} (see \cite{alwwz} for details) the equations
\bea\label{ds13}
0&=&\prt_x\Big(-V'(x)w(x)+  w^2(x) +\a w(x,x) \Big) -\frac{1}{g}, \\
&& ~ \nonumber\\
0&=&\prt_x \Big([-V'(x)+2w(x)]w(x,y)+\a w(x,x,y)\Big)+ \nonumber\\
&&\prt_y\Big([-V'(y)+2w(y)]w(x,y)+\a w(x,y,y) \Big)+\label{ds15}\\ 
&& +2 \prt_x\prt_y \Big(\frac{w(x)\mi w(y)}{x-y}\Big), \nonumber\\
&& ~ \nonumber\\
0&=&\prt_x \Big([-V'(x)+2w(x)]w(x,y,z) + \a w(x,x,y,z)\Big)+ \nonumber\\
&&  \prt_y \Big([-V'(y)+2w(y)]w(x,y,z) + \a w(x,y,y,z)\Big)+ \nonumber\\
&&  \prt_z \Big([-V'(z)+2w(z)]w(x,y,z) + \a w(x,y,z,z)\Big)+ 
\label{dsxyz} \\
&&2\prt_x [w(x,y)w(x,z)] + 2 \prt_y[ w(x,y) w(y,z)]+
2 \prt_z[ w(x,z) w(y,z)]+\nonumber\\
&&2 \left(\prt_x\prt_y \frac{w(x,z) \mi w(y,z)}{x-y} 
\pl\prt_x\prt_z \frac{w(x,y)\mi w(y,z)}{x-z}
\pl\prt_y\prt_z \frac{w(x,y) \mi w(x,z)}{y-z}\right).
\nonumber
\eea
Let us introduce the expansion\footnote{Note that both $w$ and $w_h$ are 
still $g$-dependent, although we
do not write the dependence explicitly here.}
\beq\label{ds12a}
w(x_1,\ldots,x_n) = 
\sum_{h=0}^\infty \a^{h} w_h(x_1,\ldots,x_n).
\eeq
As shown in \cite{alwwz}, $h$ can be interpreted
as the number of handles of the worldsheet, and 
the equations above can be solved iteratively in $h$.
More precisely, the equations at order $\a^0$ allow us 
to determine $w_0(x)$, $w_0(x,y)$, ..., and similarly
the equations at general order $\a^{h}$ determine $w_h(x)$,
$w_h(x,y)$, etc. For example, one finds
\beq\label{3.9}
w_0(x) = \oh\Big(V'(x)+\frac{1}{g} (x-c)\sqrt{(x-c_-)(x-c_+)}\Big),
\eeq
\beq\label{ds16}
w_0(x,y)= \frac{1}{2} \frac{1}{(x-y)^2}\left( 
\frac{xy- \oh (c_- +c_+)(x+y)+c_-c_+}{\sqrt{(x-c_-)(x-c_+)}
\sqrt{(y-c_-)(y-c_+)}} 
-1\right),
\eeq
where the constants $c$, $c_\pm$ are determined by
\beq\label{ds16a}
c^3-\l c +g=0,~~~c_\pm = -c \pm \sqrt{2g/c}.
\eeq
Writing the amplitudes in this fashion leads one to the surprising
realization that $w_0(x)$ and $w_0(x,y)$
coincide with the large-$N$ limit of the 
resolvent and the planar loop-loop correlator \cite{davidloop,ajm,am}
of the Hermitian matrix model with potential 
\beq\label{xx1}
V(M) = \frac{\l}{g} M - \frac{1}{3g} M^3\, !
\eeq
This is a potentially exciting result, because so far no 
standard formulation in terms of
matrix models has been found for a CDT model, in contrast to the ``old"
Euclidean DT models. We will in the following section 
prove a more general result, which will identify the Dyson-Schwinger 
equations derived above with the loop equations of a Hermitian
matrix model with the cubic potential \rf{xx1}.

\section{Matrix loop equations}

Let $M$ denote an $N\times N$ Hermitian matrix, $V(M)$ a potential of
the form
\beq\label{yy10}
V(M) = -\sum_{k=1}^\infty \frac{g_k}{k} \, M^k,
\eeq
and define the functions
\beq\label{yy11}
W(x_1,\ldots,x_n) = N^{n-2} 
\left\la (\tr \frac{1}{x_1-M})\cdots (\tr \frac{1}{x_1-M})
\right\ra_{c}.
\eeq
The subscript $c$ in $\la O_1(M)\cdots O_n(M) \ra_c$ 
denotes the connected part of the expectation value,
which itself is defined as
\beq\label{xx3}
\la O_1(M)\cdots O_n(M) \ra = 
\frac{\int \d M \; O_1(M)\cdots O_n(M)\;\e^{-N \tr V(M)}}{\int 
\d M \; \e^{-N \tr V(M)}}.
\eeq
It is well known that the matrix integrals corresponding to \rf{yy11} 
possess a large-$N$ expansion. 
Assume we have the so-called one-cut solution related to this 
expansion. The invariance of the matrix integral under a
change in variables leads to the loop equation \cite{davidloop,ajm,am}
\beq\label{yy1}
\int_C \frac{\d z}{2\pi i} \; \frac{V'(z)}{x-z}\; W(z) = W^2(x) + 
\frac{1}{N^2}\, W(x,x),
\eeq  
where the integration contour $C$ encloses the cut,
but not the point $x$. From this equation one can obtain
the equations for the multi-loop correlators by differentiating
with respect to the coupling constants $g_k$ in terms of the so-called 
loop insertion operator \cite{ajm,ackm} according to 
\beq\label{yy2}
W(x_1,\ldots,x_n) = \frac{\d^{n-1}}{\d V(x_2)\cdots \d V(x_n)} \; W(x_1),
\eeq
where the insertion operator is given by
\beq\label{yy3}
\frac{\d}{\d V(x)} = \sum_{k=1}^\infty \frac{k}{x^{k+1}} \; 
\frac{\d }{\d g_k}. 
\eeq
For a given potential with fixed coupling constants $g_k^0$ one uses 
these relations in the following way. Assume that $g_k$ can vary, 
act with
the loop insertion operator on \rf{yy1} as many times as needed, and then
set $g_k = g_k^0$. This leads to the desired loop equations. 
In order to compare with the Dyson-Schwinger
equations of our string field theory, we
differentiate the equations obtained with respect to $x$, and finally
find for the potential \rf{xx1} the equations
\bea\label{yy4}
0&=&{\prt_x} \Big(-V'(x)W(x) +W^2(x)\
+\frac{1}{N^2}W(x,x)\Big )-\frac{1}{g},
 \\
0&=& {\prt_x}\Big( [-V'(x)+2W(x)] W(x,y)
 +\frac{1}{N^2} W(x,x,y)\Big)+ \nonumber\\
&& + \prt_{x}\prt_y
\Big(\frac{W(x)-W(y)}{x-y}\Big),\label{yy5}\\
&& \nonumber\\
0&=& {\prt_x}\Big( [-V'(x)+2W(x)] W(x,y,z)
+\frac{1}{N^2} W(x,x,y,z) \Big)
+ \nonumber\\
&&2\prt_x\Big(W(x,z)W(x,y)\Big)+ \label{yy6}\\
&&{\prt_x \prt_y}\Big( \frac{W(x,z)-W(y,z)}{x-y}\Big)+{\prt_x \prt_z} 
\Big(\frac{W(x,y)-W(z,y)}{x-z}\Big).\nonumber
\eea
Using that $W(x_1,\ldots,x_n)$ is a symmetric function of its arguments,
we see that eqs.\ \rf{yy4}-\rf{yy6} lead to exactly the same 
coupled equations for $W$ as do \rf{ds13}-\rf{dsxyz} for $w$ {\it if} we identify
\beq\label{N}
\a = \frac{1}{N^2}.
\eeq
In this case
the discussion surrounding the expansion \rf{ds12a} is 
nothing but the standard discussion of the large-$N$ expansion  
\beq\label{yy7}
W(x_1,\ldots,x_n) = \sum_{h=0}^\infty \frac{1}{N^{2h}} 
\;W_h(x_1,\ldots,x_n)
\eeq 
of the multi-loop correlators (see, for instance, \cite{ackm} or 
the more recent papers \cite{eynard,ce}). The iterative solution
of these so-called loop equations is uniquely determined by
$W_0(x)$ (and the assumption that $W(x_1,\ldots,x_n)$ is 
analytic in those $x_i$ that do not belong to the cut of the matrix model),
and we have already seen that $W_0(x)=w_0(x)$.

\section{Discussion and Outlook}

Let us consider the matrix model corresponding to the potential
\rf{xx1}. We can perform a simple change of variables 
$M \to -M -\sla$ in the matrix integral
to obtain a standard matrix integral
\beq\label{zz1}
Z(m,g) = \int \d M \; \e^{-N V(M)},
\eeq
where the new potential (up to an irrelevant constant term) is given by
\beq\label{zz2}
V(M) = \frac{1}{g}\; \Big(\oh m M^2 +\frac{1}{3}{M^3}\Big),~~~m=2\sqrt{\l}.
\eeq
It is amusing to note that the matrix integral \rf{zz1} 
is precisely the kind of matrix integral considered  
in the so-called Dijkgraaf-Vafa correspondence \cite{dv}, where $V(\Phi)$ is the 
tree-level superpotential of the adjoint 
chiral field $\Phi$, which breaks the 
supersymmetry of the unitary gauge theory from $\cN = 2$ to $\cN = 1$.   
If one demands that this tree-level potential correspond to a 
renormalizable theory, its form is
essentially unique, and precisely of the form \rf{zz2} originally 
used by Dijkgraaf and Vafa, with $g$ a 
dimension-three coupling constant coming from topological string theory 
and in the DV-correspondence related to the glueball superfield condensate in
the gauge theory. 

In the ``old'' matrix model representation of non-critical strings
and 2d gravity one had to perform a fine-tuning of the coupling 
constants in order to obtain a continuum string or quantum 
gravity theory. This implemented the 
gluing of triangles (or, more generally, of squares, pentagons, etc.)
which served as a regularization of the worldsheet. The fine-tuning
of the coupling constants reflected the fact that the link length of the 
triangles (the lattice spacing of the dynamical lattice)
was taken to zero in the continuum limit. 
The situation here is different. Although
CDT can be constructively defined as the continuum limit of a dynamical
lattice, we have in the present work been dealing only
with the associated continuum theory.
Thus in our case the matrix model with the potential \rf{xx1} (or \rf{zz1}) 
already describes a {\it continuum} theory of 2d quantum gravity. Its 
coupling constants can be viewed as continuum coupling constants
and the role of $N$ is exactly as in the original context of 
QCD, namely, to {\it reorganize} the expansion in the coupling 
constant $g$. 
't Hooft's large-$N$ expansion of QCD is a reorganization of the perturbative
series in the Yang-Mills coupling $g_{\rm{YM}}$, with $1/N$ taking the role
of a new expansion parameter. In this framework,
after the coefficient of the term $1/N^{2h}$ of some 
observable has been calculated as function of the 't Hooft coupling 
$ g_H^2= g_{\rm{YM}}^2 N$, one must take $N=3$ for $SU(3)$, say.
The situation in CDT string field theory is entirely analogous: starting from a perturbative expansion
in the ``string coupling constant'' $g$ (in fact, in the dimensionless
coupling constant $g/\l^{3/2}$, as described in \cite{alwz,alwwz}),
we can reorganize it
as a topological expansion in the genus of the worldsheet by 
introducing the expansion parameter $\a$. 
For the
multi-loop correlators this expansion is exactly the large-$N$ expansion
of the matrix model \rf{xx1} and the coefficients, the 
functions $W_h(x_1,\ldots,x_n)$, are exactly the multi-loop 
correlators for genus-$h$ worldsheets of the CDT string field theory
with $\a=1$.

As a ``bonus" for our treatment of generalized (and therefore slightly 
causality-violating) geometries, we also obtain a matrix formulation
of the original two-dimensional CDT model proposed in \cite{al}, where 
the spatial universe 
was {\it not} allowed to split. Working out the limit as $g \to 0$
of the various expressions derived above, we see that 
this model corresponds to the large-$N$ limit of the matrix model where
the coupling constants go to infinity, but at the same time the cut
shrinks to a point in such a way that the resolvent 
(or disk amplitude) survives, that is,
\beq\label{disk}
w_0(x) \to \frac{1}{x +\sla} = w_{CDT}(x).
\eeq     
The existence of a matrix model describing the algebraic structure of the Dyson-Schwinger
equations leads automatically to the existence of Virasoro-like operators $L_n$, $n \geq -1$
\cite{davidloop,ajm}, which can be related to redefinitions of the time variable $t$ in 
the string field theory. This 
line of reasoning has already been pursued by Ishibashi, Kawai and collaborators in the
context of non-critical string field theory. It would be interesting to perform the same analysis 
in the CDT model and show that  reparametrization under the change of time-variable
will reappear in a natural way in the model via the operators $L_n$. The results should
be simpler and more transparent than the corresponding results in non-critical string field 
theory since we have a non-trivial free Hamiltonian $H_0$ in the CDT model.

\section*{Acknowledgment}
JA, RL, WW and SZ acknowledge support by
ENRAGE (European Network on
Random Geometry), a Marie Curie Research Training Network in the
European Community's Sixth Framework Programme, network contract
MRTN-CT-2004-005616. RL acknowledges
support by the Netherlands
Organisation for Scientific Research (NWO) under their VICI
program.

\end{document}